\documentclass[preprint,showpacs,preprintnumbers,amsmath,amssymb]{revtex4}
\usepackage{graphicx}
\usepackage{dcolumn}
\usepackage{bm}
\usepackage{epsfig}
\begin{document}
\title{The thermodynamic properties of weakly interacting quark gluon plasma
via the one gluon exchange interaction}
\author{M. Modarres}\altaffiliation {Corresponding author, Email:
mmodares@ut.ac.ir, Fax:+98-21-61118645}
\author{A. Mohamadnejad}\altaffiliation {Email: a.mohamadnejad@ut.ac.ir } \affiliation{Physics Department, University of
Tehran, North-Kargar Ave., 1439955961 Tehran, Iran.}
\begin{abstract}
The thermodynamic properties of  the {\it quark gluon plasma}
($QGP$) as well as its phase diagram are calculated as a function of
baryon density (chemical potential) and temperature. The $QGP$ is
assumed to be composed of the light quarks only, i.e. the up and the
down quarks, which interact weakly and the  gluons which  are
treated as they are free. The interaction between quarks is
considered in the framework of the {\it one gluon exchange} model
which is  obtained from the Fermi liquid picture. The bag model is
used, with fixed bag pressure (${\cal B}$) for the  {\it
nonperturbative} part and the {\it quantum chromodynamics} ($QCD$)
coupling is assumed to be constant i.e. no dependence on the
temperature or the baryon density. The effect of weakly interacting
quarks on the $QGP$ phase diagram are shown and discussed. It is
demonstrated that the {\it one gluon exchange} interaction for the
massless quarks has considerable effect on the $QGP$ phase diagram
and it causes the system to reach to the confined phase at the
smaller baryon densities and temperatures. The pressure of excluded
volume hadron gas model is also used to find the transition
phase-diagram. Our results depend on the values of bag pressure and
the $QCD$ coupling constant which the latter does not have a
dramatic effect on our calculations . Finally, we compare our
results with the thermodynamic properties of strange quark matter
and the lattice $QCD$ prediction for the $QGP$ transition critical
temperature.
\end{abstract}
\pacs{24.85.+p, 12.38.Mh, 12.39.Ba, 25.75.-q \\ Keywords: quark
gluon plasma, one gluon exchange, Fermi liquid model} \maketitle
\section{\label{sec:level1}Introduction}
The  {\it quark gluon plasma} ($QGP$) usually is defined as  the
phase of {\it Quantum Chromodynamics} ($QCD$) in which the quarks
and gluons degrees of freedom, that is normally   confined within
the hadrons, are mostly liberated. The possible phases of the $QCD$
and the precise locations of critical boundaries or   points are
currently being actively studied. In fact, revealing the $QCD$ phase
transition structure is one of the central aims of the ongoing and
the future theoretical and the experimental research in the field of
the hot and/or the dense $QCD$ \cite{Stephanov,Zakout,Kapusta,Yagi}.
It is about thirty years since the study of the hot and the dense
nuclear matter in the form of the $QGP$ has been started. The
experiments at the  $CERN$'s {\it super proton synchrotron} ($SPS$)
first tried to create the $QGP$ in the 1980s and 1990s: The first
hints of the formation of a new state of matter was obtained from
the $SPS$ data in terms of the global observable, the event-by-event
fluctuations, the direct photons and the di-leptons. The current
experiments at the $Brookhaven$ national laboratory's {\it
relativistic heavy ion collider} ($RHIC$) are still continuing these
efforts. In the April 2005, the formation of the quark matter was
tentatively confirmed by the results obtained at the $RHIC$ . The
consensus of the four $RHIC$ research groups was in favor of the
creation of the quark-gluon liquid at the very low viscosity
\cite{Arsene,Back,Adams,Adcox}.

Since,  in this new phase of matter, the  quarks and the gluons are
in the asymptotic freedom region,one expects that they interact
weakly. So, the  {\it perturbative} methods can be used for such a
system. The asymptotic freedom suggests two procedures   for the
creation of the $QGP$: $($i$)$ The recipe for the $QGP$ at high
temperature. If one treats the quarks and the gluons as the massless
noninteracting gas of molecules, such that the baryon density
vanishes, the critical temperature which above that, the hadronic
system dissolves into a system of quarks and gluons ($QGP$) is
${\cal T}_{c}=140 MeV $ \cite{Wong}. However, the modern lattice
$QCD$ calculation estimates the critical temperature, ${\cal
T}_{c}$, to be about $170$ $MeV$ \cite{JHEP} $($ii$)$ The recipe for
the $QGP$ at high baryon density. At zero temperature, the critical
baryon density required the transition to take place in $
n_{B}\simeq 0.7 fm^{-3} $ \cite{Wong} i.e. four times the empirical
nuclear matter density. On these grounds, one should expect to find
the $QGP$ in two places in the nature: Firstly, in the early
universe, about $10 ^{-5}s $ after the {\it cosmic big bang} or
secondly, at the core of {\it super-dense stars} such as the neutron
and the quark stars. This new phase of matter can also be  created
in the initial stage of the {\it little big bang} by means of the
relativistic nucleus-nucleus collisions in the heavy ion
accelerators \cite{Arsene,Back,Adams,Adcox}.

The critical temperature (critical baryon density) at the zero
baryon density (zero temperature) has been obtained simply for the
non-interacting, massless up and down quarks and gluons in the
reference \cite{Wong}. Some primary works in the zero baryon density
and in the framework of the bag model have been also presented in
the reference \cite{dey} (and the reference therein). For the more
complicated field theoretical  approaches see the references
\cite{new1,new2} and the references therein. So it would be
interesting to perform a similar calculation to reference
\cite{Wong}, but with the non zero interaction, for the region in
which both ${\cal T}\neq 0$ and $n_B\neq 0$.  So, in this work we
generalize above calculations for the region with the finite
temperature and the baryon density by considering  the weakly
interacting quarks in the framework of the {\it one gluon exchange}
scheme. Since it is intended to use the $perturbative$ method in our
calculation, it is assumed that $\alpha_c <1$. The dependence of  $
\alpha_{c} $ on the temperature and the baryon density is ignored
\cite{Modarres,Modarres1}. We perform calculations beyond the zero
hadronic pressure approximation of reference \cite{Wong} and the
pressure of the excluded volume hadron \cite{E1,E2,E3} model is also
taken into account to find the corresponding $QGP$ transition phase
diagram.

So, the paper is organized  as follows. In the section $II$ we
review the derivation of the {\it one gluon exchange} interaction
formulas \cite{Tatsumi} based on the  Landau Fermi liquid picture
\cite{Baym,Book}. The section $III$ is devoted   to the calculation
of the thermodynamics properties of the $QGP$. The result and its
comparison with the strange quark matter \cite{Modarres} are given
in the section $IV$. Finally a conclusion and summary are presented
in the section $V$.
\section{The Landau Fermi Liquid model and the one gluon exchange interaction}
The Landau Fermi liquid theory describes the  relativistic systems
such as: the nuclear matter under the extreme conditions, the quark
matter, the quark gluon plasma, and other relativistic plasmas. The
basic framework of the Landau  theory of relativistic Fermi liquids
is given by $Baym$ and $Chin$ \cite{Baym}. In this framework, one
can evaluate the energy density of a weakly interacting quarks in
$QGP$ by the following formulas:
\begin{equation}
{\cal E}={\cal E}_{kin}+{\cal E}_{pot} ,
\end{equation}
with
\begin{eqnarray}
{\cal E}_{kin}&=&\int\varepsilon_{p} n_{p}
\frac{g_{q}d^{3}p}{(2\pi)^{3}},\\ {\cal E}_{pot}&=&\frac{1}{2}\int
f_{p,k}^{unpol} n_{p} n_{k} \frac{g_{q}d^{3}p}{(2\pi)^{3}}
\frac{g_{q}d^{3}k}{(2\pi)^{3}},\label{Tatsumi}
\end{eqnarray}
where $ g_{q}={g_{spin}}\times{g_{color}}=2\times3=6 $, is the
degeneracy of quarks. $ \varepsilon_{kin} $, $ \varepsilon_{pot}$
and $ n_{i} $ are the kinetic and the potential energies and the
familiar Fermi-Dirac liquid distributions of our quasi-particles,
respectively. In the equation (\ref{Tatsumi}), $ f_{p,k}^{unpol} $
is the Landau-Fermi interaction function,  which is a criterion for
the interaction between the two quarks (the "$unpol$" superscript
refers to the unpolarized quark matter \cite{Modarres,Modarres1}).
It is related to the two-particle forward scattering amplitude i.e.:
\begin{equation}
 f_{p,k}^{unpol}
=\frac{m_q}{\varepsilon_p}\frac{m_q}{\varepsilon_k}{\cal M}_{p,k}.
\label{tatsumi2}
\end{equation}
$ {\cal M}_{p,k} $ is the commonly defined Lorentz invariant matrix
element. Now,  using  the  {\it Feynman rules} for the $QCD$,
 ${\cal M}_{p,k}$ can be calculated. Since, the direct term is proportional
to the trace of the {\it Gelman matrices}, the color symmetric
matrix element is only given by exchange contribution:
\begin{eqnarray}
{\cal M}_{p,k}= -{\cal G}^2\frac{1}{9}{\rm tr}({\lambda_a\over
2}{\lambda_a\over 2})\times& & \nonumber\\ && \hspace{-30mm} \bar
u(k)\gamma_\mu u(p)\bar u( p)\gamma^\mu u(k)\frac{-1}{(k-p)^2},
\label{tatsumi3}
\end{eqnarray}
where ${\cal G}^2=4\pi \alpha_c$ (our choice is $\alpha_c=0.2$).
Since our system is unpolarized, it is possible to sum over all the
spin states and  get the average of this quantity as:
\begin{equation}
\overline{\cal M}_{p,k}=\frac{2}{9} \frac{{\cal G}^2}{m_{q}^2}
\frac{2 m_{q}^2 - k.p}{(k-p)^2}.
\end{equation}
Then for massless up and down quarks,
\begin{equation}
 f_{p,k}^{unpol}=\frac{{\cal G}^2}{9}  \frac{1}{|p||k|}.
\end{equation}
So,  the total energy of interacting quarks can be evaluated by
using the  equations (1), (2), (3) and (7). In the next section, the
various thermodynamic quantities of the interacting $QGP$ like the
internal energy, the pressure, the entropy, etc are calculated.
\section{The thermodynamic properties of the $QGP$}
We begin with calculation of the  free ultra-relativistic massless
quarks partition function density, which follows by its
non-interacting gluon contribution's. Since we are interested in the
heavy ion processes,  it is  assumed that the initial state to be
the nuclear  matter with the equal neutron and proton densities i.e.
$n_{p}=n_{n}$. So,
\begin{equation}
n_{u}=n_{d},
\end{equation}
and the contributions of $u$  and $d$ quarks are equal. Then write
the $Fermionic$ partition  function for each quarks is written as
(${\cal V}$ is the volume),
\begin{equation}
\frac{ln{{\cal Z}_{q}}}{{\cal
V}}=\int[{ln(1+e^{-\beta(k-\mu)}})+{ln(1+e^{-\beta(k+\mu)}})]\frac{4\pi
g_{q}}{(2\pi)^{3}} k^{2} dk.
\end{equation}
This integral can be evaluated   analytically i.e.,
\begin{equation}
\frac{ln{\cal Z}_{q}}{{\cal V}}=\frac{4\pi g_{q}}{3(2\pi)^{3}}
\beta^{-3} (\frac{7\pi^{4}}{60} + \frac{\pi^{2}}{2}\nu^{2} +
\frac{1}{4} \nu^{4}),
\end{equation}
where $ \nu=\mu\beta $ and $ \beta=\frac{1}{\cal T} $ (in the
$Boltzman$ factor unit).

\begin{figure}[ht]
\includegraphics*[width=11cm]{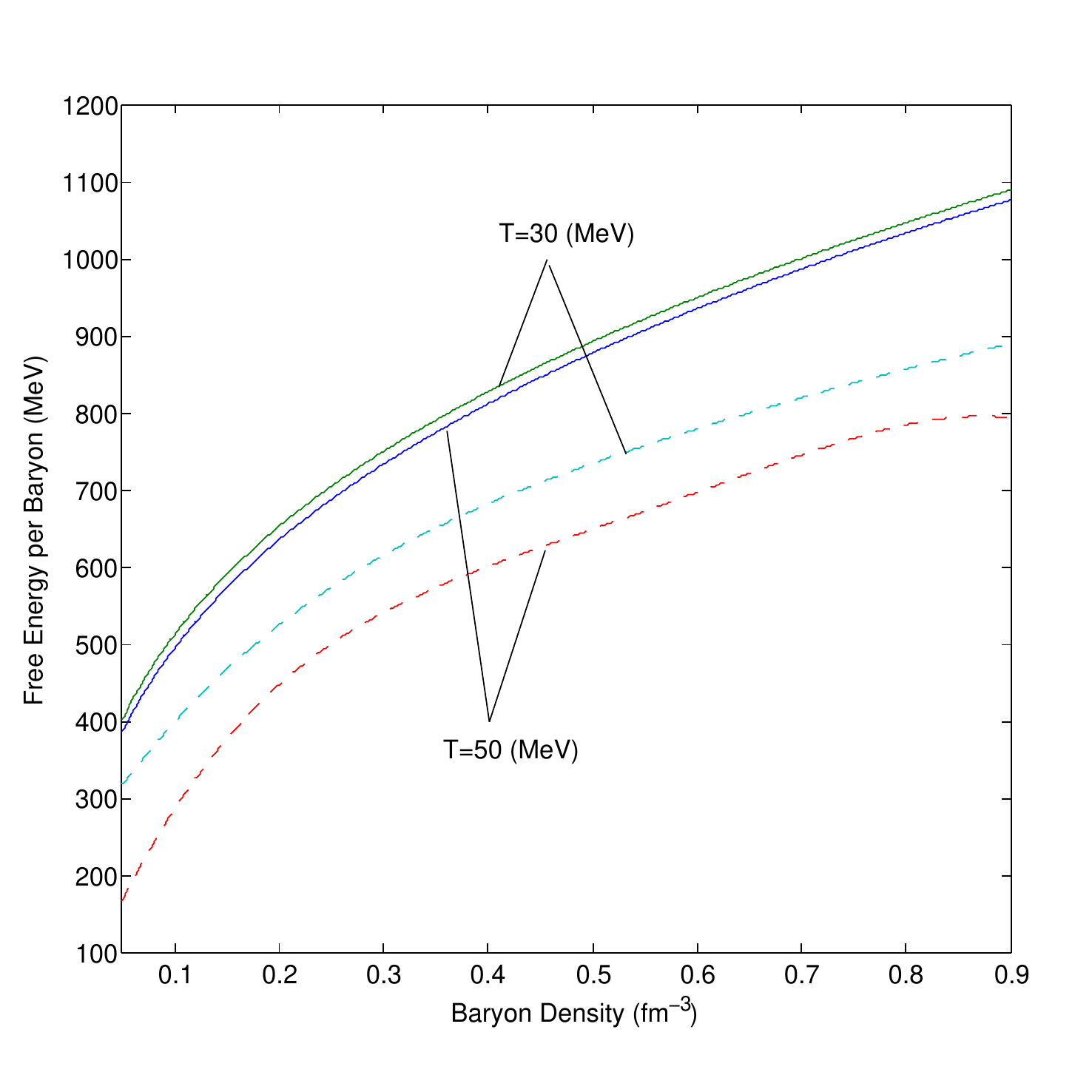}\hspace{1cm}
\caption{The density dependence of the free energy per baryon at two
different temperatures. The solid and dash curves are for the weakly
interacting $QGP$ and the strange quark matter, respectively.}
\label{Figure_1}
\end{figure}

\begin{figure}[ht]
\includegraphics*[width=11cm]{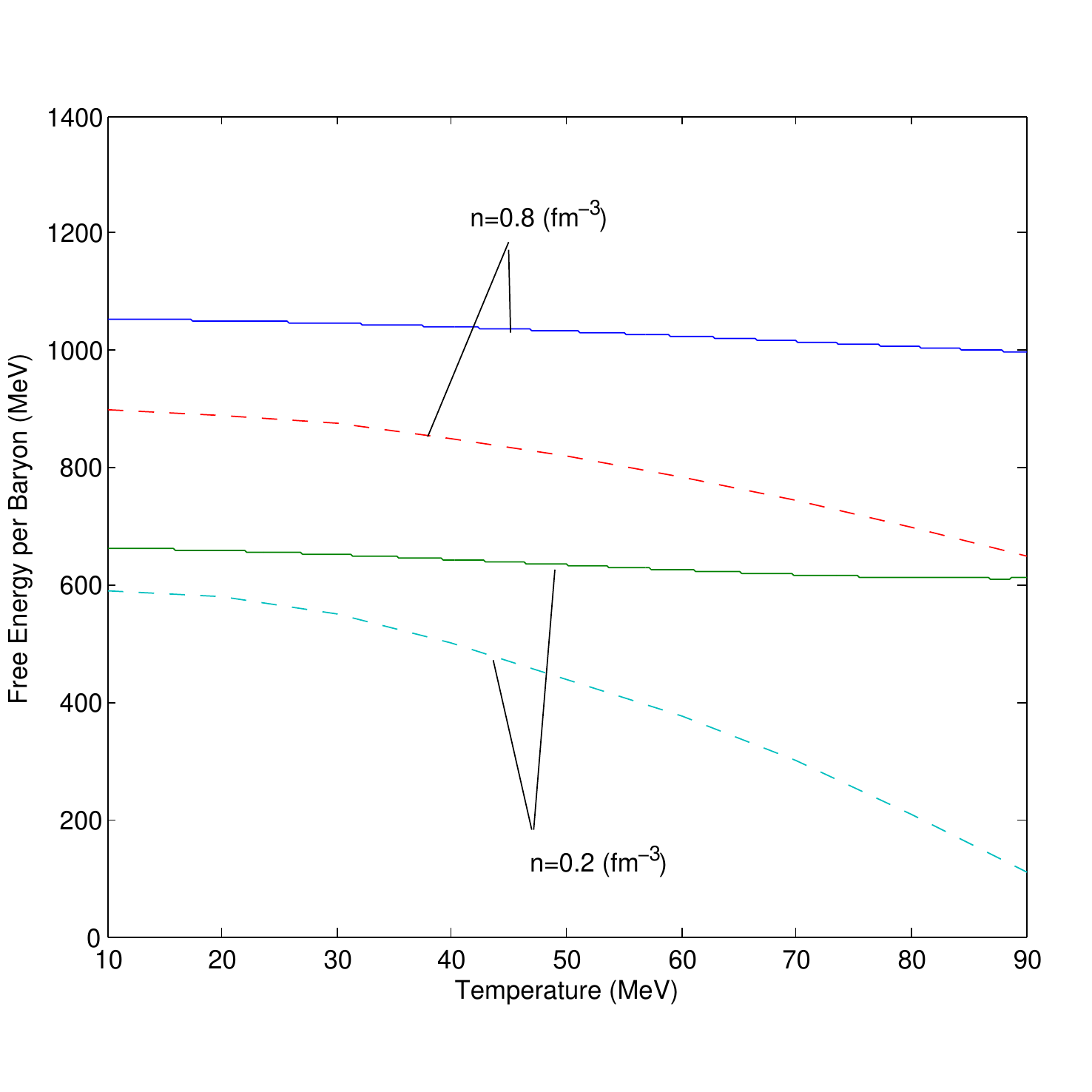}\hspace{1cm}
\caption{The temperature dependence of the free energy per baryon at
two different baryon densities. The solid and dash curves are for
the weakly interacting $QGP$ and the strange quark matter,
respectively.} \label{Figure_2}
\end{figure}

\begin{figure}[ht]
\includegraphics*[width=11cm]{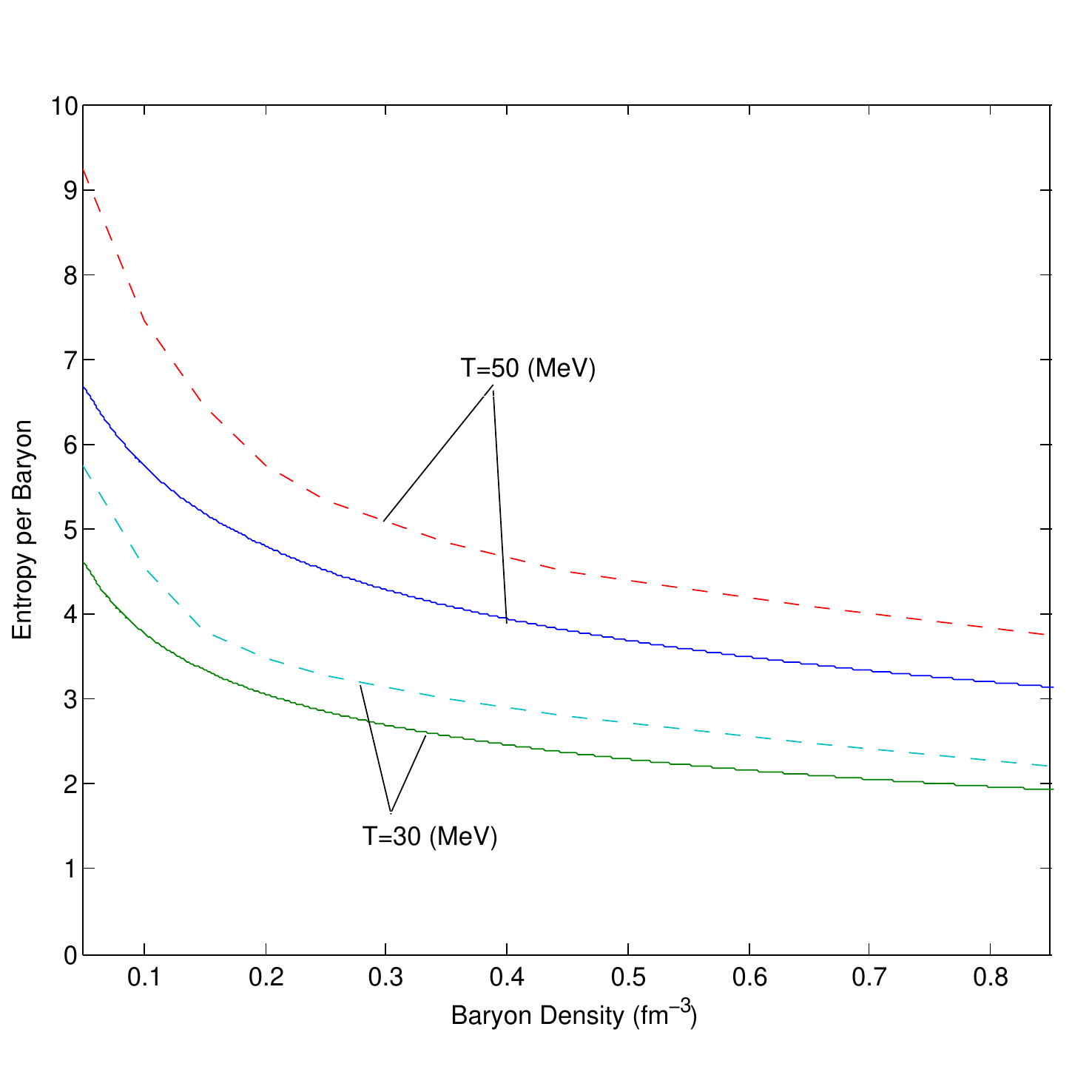}\hspace{1cm}
\caption{The density dependence of the entropy per baryon at two
different temperatures. The solid and dash curves are for the weakly
interacting $QGP$ and the strange quark matter, respectively.}
\label{Figure_3}
\end{figure}

\begin{figure}[ht]
\includegraphics*[width=11cm]{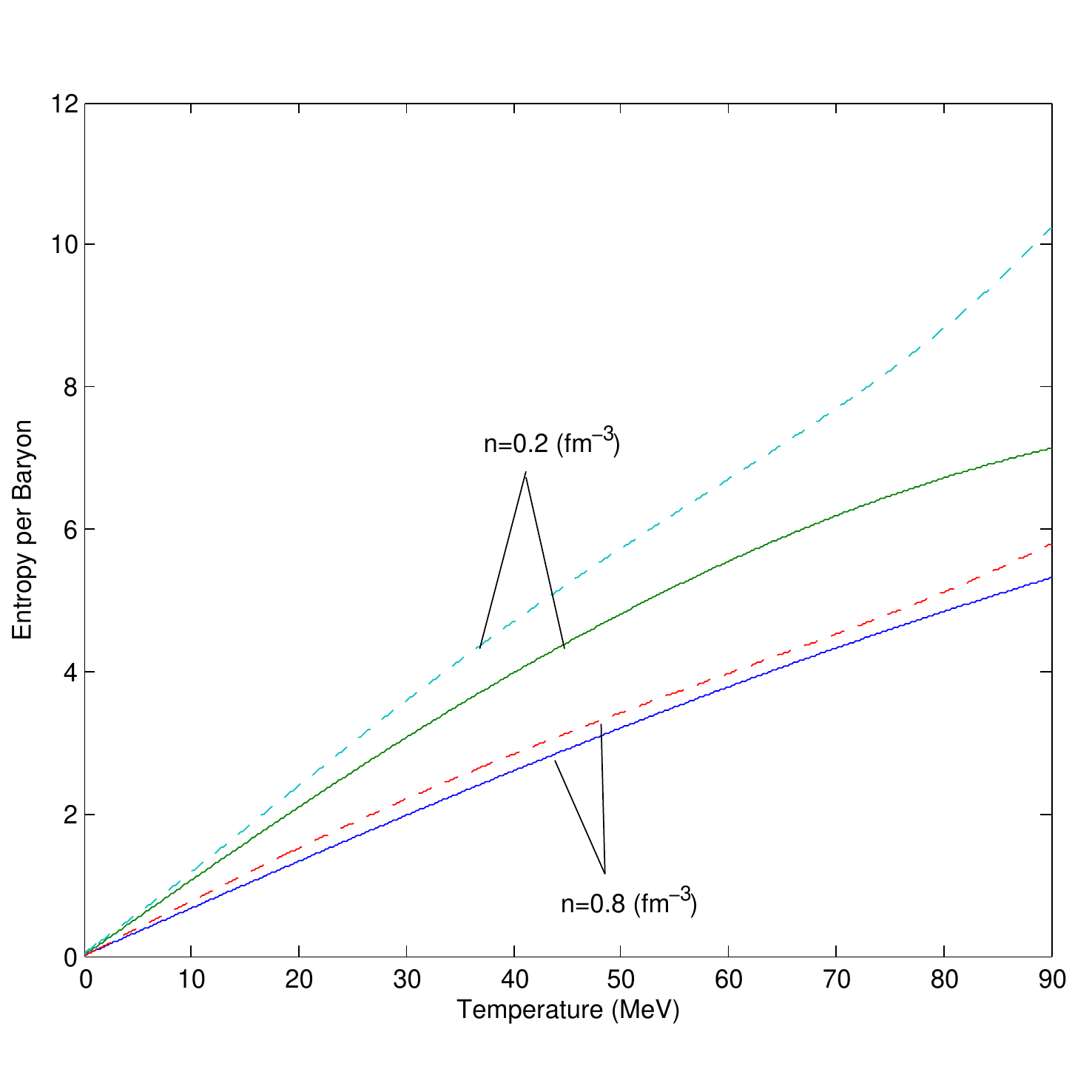}\hspace{1cm}
\caption{The entropy per baryon as a function of temperature at two
different baryon densities. The solid and dash curves are for the
weakly interacting $QGP$ and the strange quark matter,
respectively.} \label{Figure_4}
\end{figure}

\begin{figure}[ht]
\includegraphics*[width=11cm]{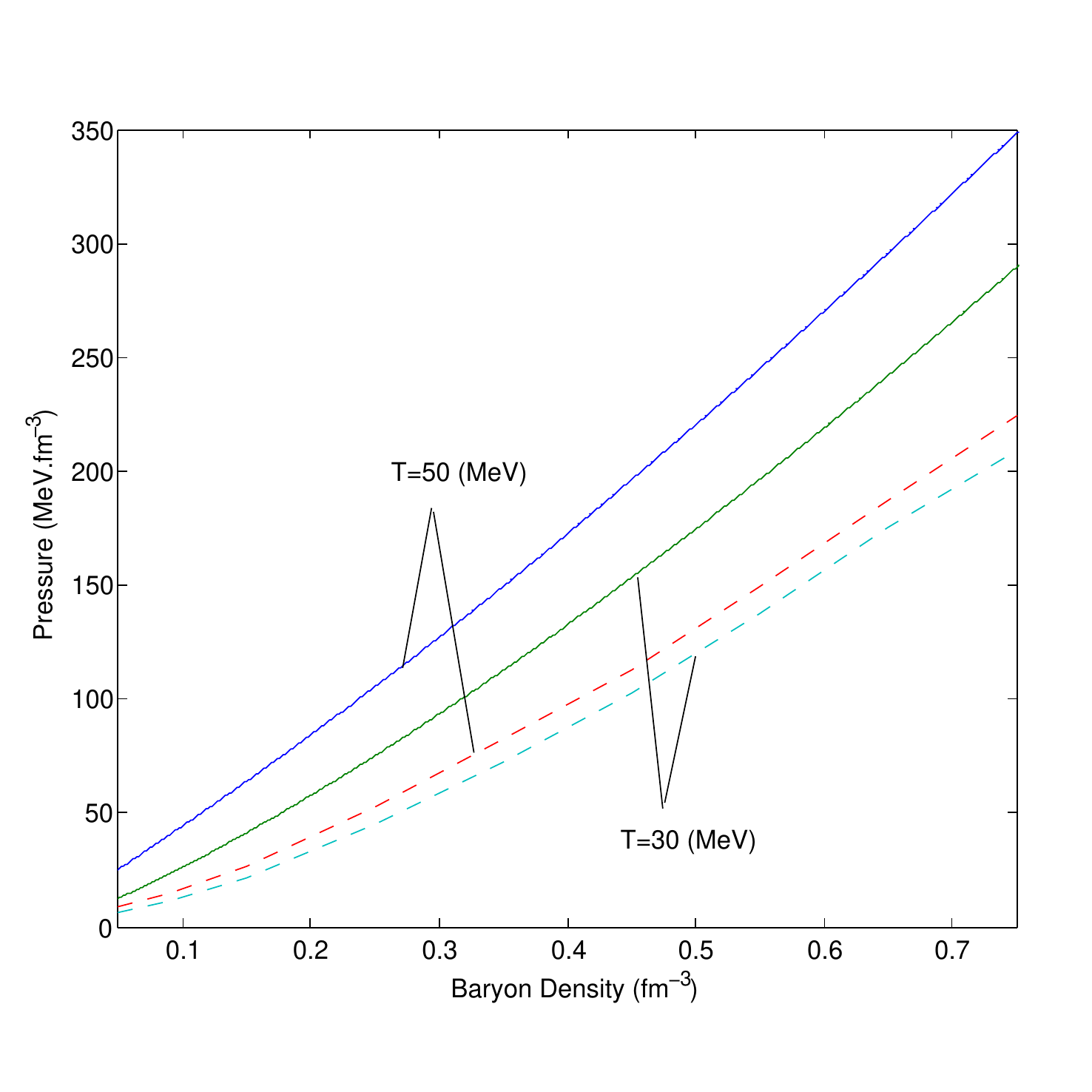}\hspace{1cm}
\caption{The $QGP$ equation of state as a function of baryon density
at two fixed temperatures. The solid and dash curves are for the
weakly interacting $QGP$ and the strange quark matter, respectively,
but without the bag constant.} \label{Figure_5}
\end{figure}

\begin{figure}[ht]
\includegraphics*[width=11cm]{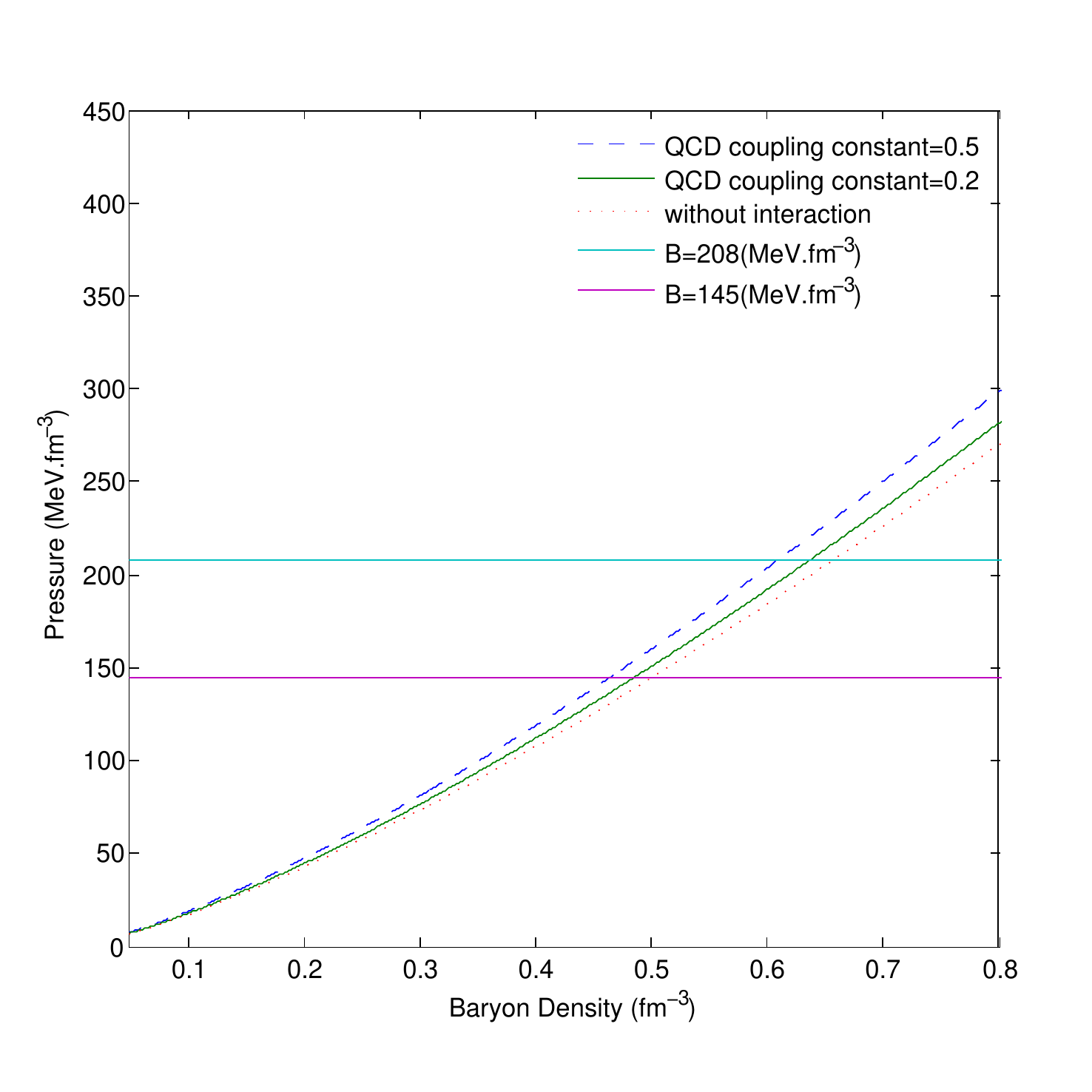}\hspace{1cm}
\caption{The pressure at zero temperature for different coupling
constants as a function of baryon density.} \label{Figure_6}
\end{figure}

\begin{figure}[ht]
\includegraphics*[width=11cm]{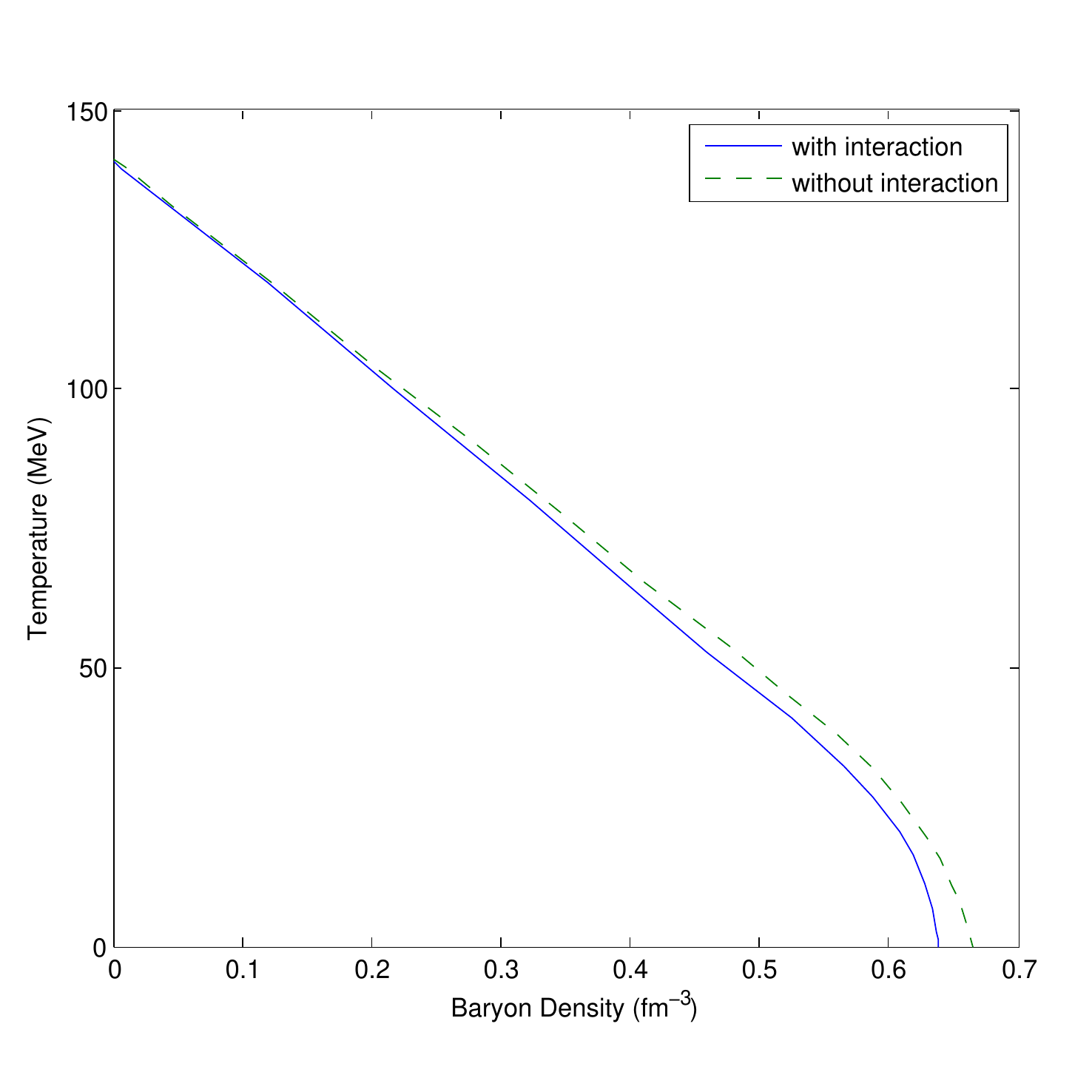}\hspace{1cm}
\caption{The phase diagram for the interacting and the
noninteracting $QGP$.} \label{Figure_7}
\end{figure}

\begin{figure}[ht]
\includegraphics*[width=11cm]{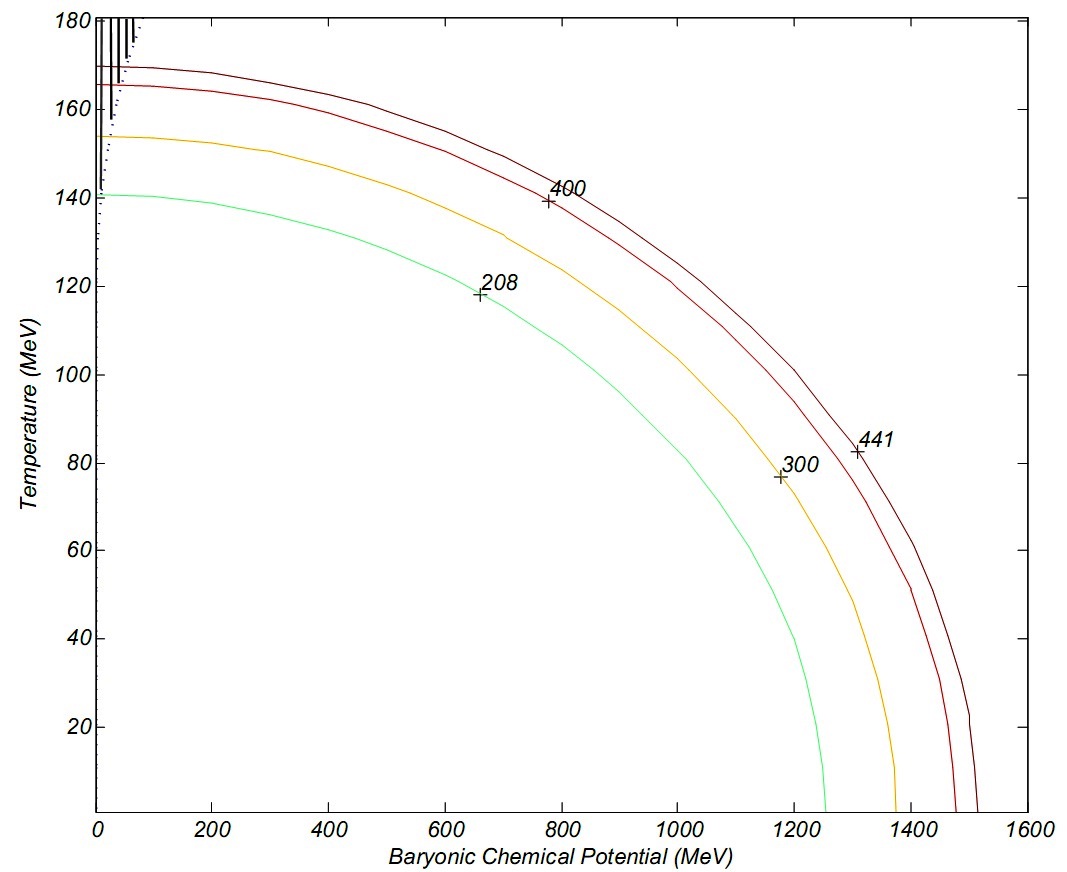}\hspace{1cm}
\caption{The phase diagram for the interacting  $QGP$ by taking into
 account the nuclear matter excluded volume pressure at four different
bag pressures ($MeV$).} \label{Figure_8}
\end{figure}

Now we are in the position to  evaluate the various thermodynamic
quantities, such as the energy density, the entropy and the free
energy from the above  partition function, especially we have,
\begin{eqnarray}
n_{q}(\mu,{\cal T})={\cal T}[\frac{\partial}{\partial \mu}
(\frac{ln{\cal Z}_{q}(\mu,{\cal T})}{{\cal
V}})]_{\cal T} & & \nonumber\\
&& \hspace{-44mm} =\frac{4\pi G_{q}}{3(2\pi)^{3}} (\pi^{2} {\cal
T}^{2} \mu+\mu^{3}),
\end{eqnarray}
where $ G_{q}=2g_q $ is the quark degeneracy for two, up and down
quarks, flavors. All of the  thermodynamic quantities are obtained
as a function of chemical potential ($ \mu $) and temperature
(${\cal T} $). The temperature is a suitable experimental quantity
but the chemical potential is not. So, it is better to rewrite the
thermodynamic quantities as a function of the temperature and the
baryon density, instead of the chemical potential. Since the baryon
number of a quark is $ \frac{1}{3} $, we have, $ n=\frac{1}{3}n_{q}
$, and:
\begin{equation}
 n=\frac{2}{3\pi^{2}} (\pi^{2} {\cal T}^{2} \mu+\mu^{3}).
\end{equation}
Then,  the chemical potential can be found as a function of $n$ and
${\cal T}$, from the above equation i.e.,
\begin{equation}
\mu(n,{\cal T})=\left( \frac{3\pi^{2}}{4}
n+\sqrt{\frac{9\pi^{4}}{16} n^{2} + \frac{\pi^{6} {\cal
T}^{6}}{27}}\right)^{\frac{1}{3}}  +  \left( \frac{3\pi^{2}}{4}
n-\sqrt{\frac{9\pi^{4}}{16} n^{2} + \frac{\pi^{6} {\cal
T}^{6}}{27}}\right)^{\frac{1}{3}},
\end{equation}
So our thermodynamic quantities become a function of baryon density
and temperature.

For the  energy density of gluons we have,
\begin{eqnarray}
\varepsilon_{g}=\int\frac{k}{e^{k/{\cal T}}-1} \frac{4\pi g_{g}
k^{2} dk}{(2\pi)^{3}} & & \nonumber\\ && \hspace{-47mm} =g_{g}
\frac{\pi^{2}}{30} {\cal T}^{4},
\end{eqnarray}
where again $ g_{g}=16 $ is the degeneracy of gluons.

Our system is ultra-relativistic, so, there is a simple relation
between the pressure and the energy density:
\begin{equation}
{\cal P}=\frac{1}{3} \varepsilon.
\end{equation}
The  bag pressure and the vacuum energy contributions should be
included, in addition to the quark and the gluon contribution
energies,
\begin{eqnarray}
\varepsilon_{QGP}(n,{\cal T})&=&\varepsilon_{q}(n,{\cal T})+\varepsilon_{g}({\cal T})+{\cal B} \\
{\cal P}_{QGP}(n,{\cal {\cal T}})&=&{\cal P}_{q}(n,{\cal T})+{\cal
P}_{g}({\cal T})-{\cal B}.
\end{eqnarray}
Having the pressure,  the entropy density of system is evaluated,
\begin{equation}
{\cal S}_{QGP}(n,{\cal T})=\left( \frac{\partial}{\partial {\cal T}}
{\cal P}_{QGP}(\mu,{\cal T})\right) _{\mu}.
\end{equation}
To study the phases of the $QGP$, one should concentrate on
  the $QGP$ and the hadronic pressures, ${\cal P}_{QGP}$  and ${\cal
P}_{Hadron}$  \cite{E1,E2,E3}, respectively, i.e. for ${\cal
P}_{QGP}<{\cal P}_{Hadron}$, the system is in the confinement phase
and the quarks and the gluons are inside the bag, but for $ {\cal
P}_{QGP}>{\cal P}_{Hadron}$, the quarks and the gluons pressures can
overcome to the bag  and the hadronic pressures and the system is in
the de-confinement phase, i.e. the $QGP$ phase is created. So,  the
phase diagram can be extracted by solving the following equation:
\begin{equation}
{\cal P}_{QGP}(n,{\cal T})={\cal P}_{q}(n,{\cal T})+{\cal
P}_{g}({\cal T})-{\cal B}={\cal P}_{Hadrons}(n(\mu),{\cal
T}),\label{N1}
\end{equation}
where in this work,  both ${\cal P}_{Hadrons}=0$ \cite{Wong} and
${\cal P}_{Hadrons}\neq 0$ \cite{E1,E2,E3} cases are considered .
For ${\cal P}_{Hadrons}\neq 0$, the excluded volume effect for the
nuclear matter equation of state is used \cite{E1,E2,E3} to
calculate the hadronic pressure (equation (29) of reference
\cite{E2}) i.e.,
\begin{equation}
{\cal P}_{Hadrons}(n(\mu),{\cal T})={\cal P}_{Hadrons}(\mu,{\cal
T})\approx {{\cal P}^{ideal}_{Hadrons}(\mu,{\cal T})\over
[1+vn(\mu)]},\label{N2}
\end{equation}
where ${\cal P}^{ideal}_{Hadrons}(\mu,{\cal T})$ is the pressure of
free ideal nucleonic (Fermion) matter \cite{E3,E4} (here, degeneracy
is 4, $m\simeq 939 MeV$, $v=4{4\over 3}\pi r^3$ and $r\simeq 1.2
fm$, e.g see the equation (9) of the reference \cite{E3}).

Now,  the interaction energy between the quarks can be calculated by
using the Landau Fermi liquid model and the results derived by us in
the previous section. The potential energy density of the
interacting massless up or down quarks  is found by using the
equations (3) and (7),
\begin{equation}
\varepsilon_{pot}(\mu,{\cal T})=\frac{1}{2}\int \frac{1}{9} {\cal
G}^{2} \frac{1}{pk} n_{p} n_{k} \frac{g_{q}d^{3}p}{(2\pi)^{3}}
\frac{g_{q}d^{3}k}{(2\pi)^{3}},
\end{equation}
where $ g_{q}=6 $ is the quark degeneracy of one flavor. By using
the Fermi-Dirac distribution and the value  of each quark flavor
potential energy i.e. the equation (21),   an analytical formula for
the total potential energy density of the two quark flavours is
written as follows:
\begin{equation}
\varepsilon_{pot}(\mu,{\cal T})=\frac{\alpha_{c} {\cal
T}^{4}}{\pi^{3}} \left( \frac{\pi^{4}}{9} + \frac{2\pi^{2}}{3}
\left( \frac{\mu}{\cal T}\right) ^{2} + \left( \frac{\mu}{\cal
T}\right) ^{4} \right).
\end{equation}
Note  that,  the potential energy for massless quarks is always
positive. So, the interaction between quarks inside the bag is
repulsive and it helps the  interacting quarks and the gluons to
penetrate from the bag more easily, rather than the noninteracting
case, and further more, the {\it one gluon exchange} interaction,
because of its repulsive properties, makes the  conditions  easier
for  the system to make the transition  to the $QGP$ phase.

The internal energy density of the $QGP$ is evaluated by performing
the summation over the interacting and the non-interacting parts of
the energy density of quarks, the vacuum energy and the gluon energy
density which were calculated before, and  having that, the other
thermodynamic quantities of the $QGP$ are found.

As it was pointed out before, it is assumed that the bag (hadronic)
pressure is the one has been used in the  reference \cite{Wong} i.e.
${\cal B}=208$$ MeV fm^{-3} $ (zero), in order to compare our phase
diagram with this reference, and since it is intended to compare our
results with the reference \cite{Modarres},  the $QCD$ coupling
constant os chosen  to be $ \alpha_{c}=0.2 $. On the other hand, for
the none zero hadronic pressure,  the baryonic chemical potential
and the bag pressure are varied to reach to the critical temperature
of the lattice $QCD$ predictions \cite{JHEP}, i.e. ${\cal T}_c=170$
$MeV$.
\section{The results of the $QGP$ calculation}
We begin by presenting the calculated    free energy per baryon for
both the $QGP$ and the strange quark matter \cite{Modarres}   in the
figure 1 (2), as a function of baryon density (temperature) at the
two different temperatures (baryon densities). The free energy for
the $QGP$ is larger than those of strange quark matter,  since we
know that the strange quark matter should  be more stable than the
$QGP$ and therefore the strange quark matter free energy should be
smaller than that of the $QGP$. As one should expect, the free
energy increases (decreases) by increasing the baryon density
(temperature). While, the $QGP$  has less temperature dependent with
respect to the strange quark matter, they have similar density
dependent at fixed temperature.

Similar comparisons are made for the entropies in the figures  3 and
4. The entropy per baryon for the $QGP$ is an increasing
(decreasing)
 function of temperature (baryon density) and  it is smaller than that of the  strange
 quark matter \cite{Modarres}.
Again their dependence on the density is the same, but they
  behave especially differently  at  larger temperatures.

The plots of  the equation of states of  both the $QGP$ (without the
effect of constant bag pressure) and the strange quark matter as a
function of baryon density at the two different temperatures are
given in the figure 5. The $QGP$ equation of state  is much harder
than that of strange quark matter at the same baryon density and
temperature.

The pressure of weakly interacting $QGP$ for the two different $QCD$
coupling constants, and the noninteracting $QGP$ (without the
effects of constant bag pressure) at zero temperature  as a function
of baryon density is shown in the figure 6. The increase in the
interaction strength makes the  pressure to rise, and therefore at
the smaller baryon densities, the pressure of quarks becomes equal
to the bag pressure. So, the interaction facilitates   the quarks
transition to  the deconfined phase at lower density.  The $QCD$
coupling constant also plays the same
  role, i.e it will reduce the transition density.

Finally, the phase diagrams for both the interacting and the
noninteracting $QGP$ are shown in the figure 7 for ${\cal
P}_{Hadrons}(n(\mu),{\cal T})=0$. The  {\it one gluon exchange}
interaction which is repulsive, causes  to get the $QGP$ at the
smaller baryon densities and temperatures. As it was pointed out
before, the reason is very simple, the repulsive interaction between
quarks helps them to escape from the bags. So the formation of the
$QGP$ happens much easier for the interacting quarks than the
noninteracting one. But, the critical temperature is about $140
Mev$, which is much less than the lattice $QCD$ suggestion of $170
MeV$. In the figure 8, the hadronic pressure has been also taken
into the account (see the equations (\ref{N1}) and the reference
\ref{N2}). The slashed area is the forbidden region, i.e. for the
bag pressure approximately   larger than $200 MeV$ there is no
critical temperature with the zero baryonic chemical potential
(density). With the bag pressure about $441 MeV$ (note that the bag
pressure estimated to as large as $500$ $MeV$ \cite{Yagi}) and the
low baryonic chemical potential (density) it is possible to get
results near to the lattice $QCD$ prediction, i.e. ${\cal T}_c=170$
$MeV$.
\section{conclusion and summary}
In conclusion, the {\it one gluon exchange} interaction was used to
evaluate the strength of potential energy of the $QGP$  in the Fermi
liquid model. By calculating the $QGP$ partition function, the
different thermodynamic properties of the $QGP$ as a function of
baryon density and temperature for the both interacting and the
non-interacting cases were discussed. It was found that the $QGP$
internal and free energies are much larger than those of strange
quark matter. On the other side, if we consider the massive quarks
like the strange quarks in our the $QGP$, the potential energy
becomes a negative quantity, but for the massless quarks it is
always a positive quantity, therefore the internal and the free
energy densities of strange quark matter become smaller than the
$QGP$ ones. We have seen how the {\it one gluon exchange}
interaction for the massless quarks affects  the phase diagram of
the $QGP$ and causes the system to reach to the deconfined phase at
the smaller baryon densities and temperatures. Our results depend on
the values of  bag pressure and the $QCD$ coupling constant which
the latter does not have a dramatic effect on our results. The
increase of the hadronic and the bag pressure can improve our
results toward the lattice $QCD$ calculations. In the future works,
we could adjust our phase diagram to get the relation between the
bag constant and the $QCD$ coupling constant. On the other hand it
is possible to generalize our method for the non-constant $QCD$
coupling and the bag pressure. Finally, we can also add the
interaction between the gluons to our present calculations.

\begin{acknowledgments}
We would like to acknowledge  the Research Council of University of
Tehran and Institute for Research and Planning in Higher Education
for the grants provided for us.
\end{acknowledgments}

\end{document}